\documentclass[journal,comsoc]{IEEEtran}

\usepackage[T1]{fontenc}
\usepackage{graphicx}

\usepackage{algorithm}
\usepackage{algpseudocode}
\usepackage{multicol}
\usepackage{float}
\usepackage{bm}
\usepackage{pifont}
\usepackage{relsize}
\usepackage[cmintegrals]{newtxmath}
\usepackage{mathrsfs}
\usepackage{amsmath}
\usepackage{subcaption}
\usepackage{multirow}

\ifCLASSINFOpdf
\else
\fi

\usepackage[cmintegrals]{newtxmath}

\usepackage{fancyhdr}

\begin{document}

\pagestyle{fancy}
\fancyhead[CE,CO]{\scriptsize{© 2020 IEEE.  Personal use of this material is permitted.  Permission from IEEE must be obtained for all other uses, in any current or future media, including reprinting/republishing this material for advertising or promotional purposes, creating new collective works, for resale or redistribution to servers or lists, or reuse of any copyrighted component of this work in other works.}}

\renewcommand{\headrulewidth}{0pt}
\renewcommand{\footrulewidth}{0pt}

\title{NOMA for Multiple Access Channel and Broadcast Channel in Indoor VLC}
\author{T. Uday,~\IEEEmembership{Student Member,~IEEE}, Abhinav Kumar,~\IEEEmembership{Senior Member,~IEEE},
       and L. Natarajan,~\IEEEmembership{Member,~IEEE}

\thanks{The authors are with the Department of Electrical Engineering, Indian Institute of Technology Hyderabad, Telangana, India.
 (e-mail: \{ee18resch11005, abhinavkumar, lakshminatarajan\}@iith.ac.in). This work was supported in part by the Department of Science and Technology (DST), Govt. of India (Ref. No. TMD/CERI/BEE/2016/059(G)).}}%
\maketitle
\thispagestyle{fancy}

\begin{abstract}
Orthogonal frequency division multiplexing (OFDM) based non-orthogonal multiple access (NOMA) has increased complexity and reduced spectral efficiency in visible light communications (VLC) NOMA compared to radio-frequency NOMA due to non-negative real value constraints on transmit symbols. To address this issue, we propose a generalized non-OFDM based scheme for two scenarios of indoor VLC; i) Multiple access channel (MAC), ii) Broadcast channel (BC).
We evaluate the performance of the proposed scheme for MAC using successive interference cancellation (SIC) based decoding, joint maximum likelihood (JML) decoding, and a combination of SIC and JML decoding. For BC, we evaluate the performance using SIC based decoding. 
It is observed that, for MAC, the proposed scheme with JML decoding performs better than the state-of-the-art orthogonal multiple access both in terms of  bit error rate (BER) and computations. For BC, the proposed scheme is computationally efficient with trade-off on BER.

\end{abstract}
\IEEEpeerreviewmaketitle

\section{Introduction}
\IEEEPARstart{L}{ight} emitting diode (LED) based indoor visible light communication (VLC) transmits data by modulating the light intensity, which is typically referred to as intensity modulation.
Recently, non-orthogonal multiple access (NOMA) technique has been proposed for VLC \cite{intro_1_2,intro_3_1,intro_3_2}. In NOMA, to decode the data at receiver (Rx), successive interference cancellation (SIC) is performed on the received power domain superposed signal \cite{intro_3_2}.
In \cite{intro_3_1,intro_3_2}, an on-off keying based implementation has been considered. However, higher order modulation schemes have not been discussed. 

Orthogonal frequency division multiplexing (OFDM) based schemes have been proposed for VLC namely direct current (DC)-biased optical OFDM (DCO-OFDM) \cite{ofdm_1,ofdm_2} and asymmetrically clipped optical OFDM (ACO-OFDM) \cite{ofdm_3}. However, these techniques require Hermitian symmetry plus inverse fast Fourier transform (IFFT) to convert the complex symbols to real domain. Then, DC biasing or clipping the negative part of the signal is done in DCO-OFDM and ACO-OFDM, respectively. Hence, the complexity involved in the implementation of these techniques is higher. It is also observed that spectral efficiency (in bits per channel use (bpcu)) of DCO-OFDM and ACO-OFDM is half that of radio frequency (RF) NOMA (RF-NOMA) due to the Hermitian symmetry used to convert the complex symbols to real domain.
\begin{figure}[!t]
\centering
\includegraphics[width=3.25in,height=1.21in]{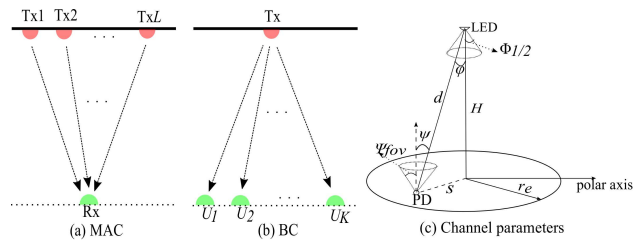}
\caption{System model.}
\label{system_model}
\end{figure}
In power domain NOMA, symbols normalized to unit power are multiplexed considering the power allocation coefficients.
In VLC, the modulation symbols can only be non-negative real values under the typical consideration that the symbols are the LED intensity levels. Due to this non-negativity constraint on the modulation symbols, the NOMA schemes proposed for RF mobile communication cannot be directly applied to VLC.
Therefore, in this letter, we propose a scheme that assigns a set of power allocation coefficients to each transmitter (Tx)/user (Tx in multiple access channel (MAC) and user in broadcast channel (BC)) based on the required spectral efficiency for that Tx/user unlike a single power allocation coefficient for each of the Txs/users in RF-NOMA. These coefficients themselves act as the modulation symbol set and avoids Hermitian symmetry, the DC biasing or the clipping of negative portion of the signal which improves spectral efficiency compared to OFDM-NOMA in VLC and makes the system implementation simpler. To the best of our knowledge, this is the first work to propose non-OFDM based NOMA for VLC with arbitrary modulation order in multiple access and broadcast channel.

To implement the proposed scheme, we consider two scenarios of indoor VLC; i) MAC, where multiple Txs are communicating to a single user, with each Tx using a single LED and receiver using a photo diode (PD)  \cite{ref_mac}. ii) BC, where a single Tx is communicating to multiple users.
In terms of bit error rate (BER) and computational complexity, we evaluate MAC using SIC based decoding, joint maximum likelihood (JML) decoding, and $M$ JML + $(L-M)$ SIC decoding, where, the data from $M$ Txs is decoded using JML decoding and the data from $(L-M)$ Txs is decoded using SIC, given a total of $L$ Txs $\left( M\geq 2 \,\, \mbox{and}\,\,M<L\right)  $. The BC is evaluated using SIC based decoding.

\textit{Notation:} We use $\lceil.\rceil$, $\lfloor.\rfloor$, $\mathbb{Z^+}$, $|.|$, $||.||$, and $\mathop{\mathbb{E}}$ for ceil operation, floor operation, set of positive integers excluding zero, absolute value, Frobenius norm, and expectation function respectively.
\section{System Model}
The system model to implement the proposed scheme for MAC and BC is shown in Fig.~\ref{system_model} (a) and Fig.~\ref{system_model} (b), respectively, with LEDs used as the Txs and PDs used as the Rx/user. Given this,
the indoor channel gain between the PD and the LED, denoted by $h$ is given as \cite{intro_1_2}
$$
h=
\begin{cases}
\frac{(\zeta+1)A_DR_p \textrm{cos}(\phi)^\zeta T(\psi)g(\psi)\textrm{cos}(\psi)}{2\pi d^2} ,\,\,\,\,\,\, \psi\in [0, \,\, \psi_{fov}]\,,\\
0,\,\,\,\,\,\,\,\,\,\,\,\,\,\,\,\,\,\,\,\,\,\,\,\,\,\,\,\,\,\,\,\,\,\,\,\,\,\,\,\,\,\,\,\,\,\,\,\,\,\,\,\,\,\,\,\,\,\,\,\,\,\,\,\,\,\,\,\,\,\,\,\,\,\,\,\,\,\, \psi>\psi_{fov}\,,\\
\end{cases}
$$
where, $\zeta$ is the order of Lambertian radiation pattern given by $\zeta=-1/\textrm{log}_2(\textrm{cos}(\Phi_{1/2}))$  such that $\Phi_{1/2}$ is the angle at half power of LED, $A_D$ denotes the detection area of the PD at the Rx, $R_p$ denotes the responsivity of the PD, $T(\psi)$ represents the gain of the optical filter used at the PD, where, $\psi$ is the angle of incidence at the PD from LED as shown in Fig.~\ref{system_model} (c), $d$ is the distance between the PD and the LED, $g(\psi)$ represents the gain of the optical concentrator, $\psi_{fov}$ is the field of view of the PD, $\phi$ angle of emission at the LED with respect to the PD, $H$ is the vertical distance from the LED to surface, $S$ is the radial distance between the PD and the LED from top view, and $r_e$ is the LED coverage distance from the top view.
\begin{figure}[!t]
\centering
\includegraphics[width=2.75in,height=1.25in]{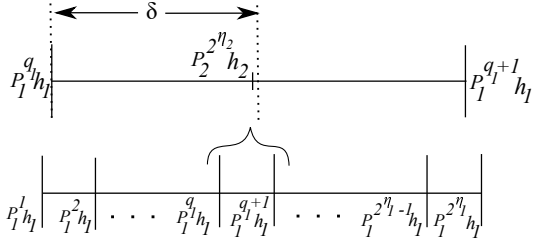}
\caption{Figure to explain zero BER in zero noise variance and perfect CSI conditions.}
\label{const_points}
\end{figure}
\section{Proposed Coding Scheme}
In this section, we first propose the scheme for MAC and then extend it to BC.
\subsection{MAC}For MAC, we assume the transmission duration of each symbol/constellation point is same for all the  Txs (LEDs) and also the transmissions are synchronized in time.
Let $L$ be the total number of Txs, $\mathit{h}_x$ be the channel gain between $x^{th}$ Tx and PD Rx. We consider that the channel gains are known at the Txs and are sorted such that 
$\mathit{h}_1 \leq  \mathit{h}_2 \leq \ldots \leq  \mathit{h}_L$. Based on the NOMA principle, higher channel gain corresponding Txs are allocated lower transmit power and vice versa \cite{intro_1_2}. Given this, the condition to perform successful SIC at the Rx is given as follows
\begin{equation}
\mathit{P}_1^{q_1}\mathit{h}_1 > \mathit{P}_2^{q_2} \mathit{h}_2 > \ldots > \mathit{P_L}^{q_L} \mathit{h}_L, 
\label{scenario1_constraint}
\end{equation}
where, $\mathit{P_x}^{q_x} \in$ $ \mathbb{Z^+}$ denotes the power transmitted by the $x^{th}$ Tx, $x \in \mathbf{X} = \{1, 2, \ldots, L\}$, and $q_x$ is the index of transmit power values assigned to $x^{th}$ Tx such that $q_x \in \mathbf{Q}_x  = \{1, 2, \ldots,2^{\eta_x}\}$. Here, $\eta_x$ denotes the spectral efficiency of the $x^{th}$ Tx. The $\mathit{P_x}^{q_x}$ are the constellation points assigned to $x^{th}$ user and to achieve a spectral efficiency of $\eta_x$, $2^{\eta_x}$ constellation points are required as given by $ \mathbf{Q}_x$. For example, for the $1^{st}$ Tx, the transmit power values are given as a set $\{ \mathit{P}_1^1, \mathit{P}_1^2, \ldots,\mathit{P}_1^{2^{\eta_1}}\}$. These transmit power values are treated as constellation points in the power domain.
\begin{algorithm}[t]
  \caption{Generate the constellation points for $L$ Txs with desired spectral efficiency for each Tx.}\label{algo_1}
  \begin{algorithmic}[1]
  \State Choose $\eta_L$ and assign integer values for the constellation points corresponding to $L^{th}$ Tx as
  $1, 2, \ldots, 2^{\eta_L}$ for $\mathit{P_L}^1, \mathit{P_L}^2, \ldots, \mathit{P_L}^{2^{\eta_L}}$, respectively.   
 \For{\texttt{$x=L-1$ to $1$}}
             \State $\mathit{P_x}^1 = 2^{\eta_{x+1}}+1$
                  \For{\texttt{$\mathit{q_x}=1$ to $2^{\eta_x}-1$}}
                  \State $   \mathit{P_x}^{q_x+1}  = \left\lfloor \Delta_x^{q_x+1}+1 \right\rfloor$, \mbox{where}, $\Delta_x^{q_x+1}$ is as in \eqref{for_algo}.
            \EndFor  
           \EndFor         
  \end{algorithmic}
\end{algorithm}
\begin{algorithm}[t]
  \caption{Decode using SIC.}\label{SIC_decoding}
  \begin{algorithmic}[1]
  \State $\hat{\mathit{P}_1} = \min_{\forall q_1 \in \mathbf{Q}_1} ||y-\widetilde{\mathit{P}_1^{q_1}} \mathit{h}_1 ||$  
 \For{\texttt{$x=2$ to $x=L$ }}
             \State $\hat{\mathit{P}_x} = \min_{\forall q_x \in \mathbf{Q}_x} ||y-\sum_{r=1}^{x-1} \hat{\mathit{P}}_{r}\mathit{h}_{r}-\widetilde{\mathit{P}_x^{q_x}} \mathit{h}_x ||$                  
           \EndFor     
  \end{algorithmic}
\end{algorithm}
We normalize the constellation points as follows 
\begin{equation*}
 \widetilde{ \mathit{P_x^{q_x}}} = \frac{\mathit{P_x^{q_x}}}{\sum_{x=1}^{L}\sum_{q=1}^{\mathit{2^{\eta_x}}}P_x^{q_x}} \,\forall\, x \in \mathbf{X}, q_x \in \mathbf{Q}_x,
\end{equation*}
where, $\widetilde{ \mathit{P_x^q}}$ is the normalized constellation point. These normalized constellation points can be scaled for brightness control such that $\sum_{x=1}^{L} \widetilde{\mathit{P_x}^{2^{\eta_x}}} \leq \mathcal{P}$, where $\mathcal{P}$ is the total available transmit power per channel use for all the Txs. The highest value of constellation point for each Tx, $\widetilde{\mathit{P_x}^{2^{\eta_x}}}$ is chosen so that in any channel use, the total transmit power will not exceed  $\mathcal{P}$. Given this, the received signal, $\mathit{y}$ is given as
$\mathit{y}=\sum_{x=1}^{L} \widetilde{\mathit{P}_x^{q_x}}\mathit{h}_x +\mathit{n},$
where, $\mathit{n}$ is the real valued additive white Gaussian noise (AWGN) with 0 mean and $\sigma^2$ variance, as in \cite{2}.

At the Rx, for decoding the data transmitted by any Tx, the power received from higher power assigned Txs is removed using SIC and the power received from remaining Txs is treated as noise.
For the proposed scheme, we assume that the distance between the consecutive constellation points assigned for individual Txs is same, and the constellation points are in the increasing order which is given as follows 
\begin{equation}
\mathit{P_x}^1 < \mathit{P_x}^2 < \ldots < \mathit{P_x}^{2^{\eta_x}}, \,\, s.t., \,\, |\mathit{P_x}^{q_x} - \mathit{P_x}^{q_x+1}| = \lambda_x,
\label{constraint_3}
\end{equation}
where, $\lambda_x$ is the distance between the consecutive constellation points assigned to the $x^{th}$ Tx and it's value can vary across Txs.

We define ideal conditions as zero noise variance ($\sigma^2=0$) along with availability of perfect channel state information (CSI) at all the Txs and Rx.
In Fig.~\ref{const_points}, the possible received power values corresponding to Tx1 in ideal conditions is shown. Considering any two consecutive received constellation points, $\mathit{P}_1^{q_1}\mathit{h}_1$ and $\mathit{P}_1^{q_1+1}\mathit{h}_1$, the possibility of non-zero BER even in ideal conditions is explained in the presence of an interferer (Tx2).
From Fig.~\ref{const_points}, in case $\mathit{P}_1^{q_1}\mathit{h}_1+\mathit{P}_2^{q_2}\mathit{h}_2$ is received and $\mathit{P}_2^{2^{\eta_2}}\mathit{h}_2 > \delta$, where, $\delta =\left(\mathit{P}_1^{q_1+1}\mathit{h}_1-\mathit{P}_1^{q_1}\mathit{h}_1\right)/2 =\left(\mathit{P}_1^{q_1+1}-\mathit{P}_1^{q_1}\right)\mathit{h}_1/2 =\lambda_1\mathit{h}_1/2$ is as shown in Fig.~\ref{const_points}. Then, at the Rx, $\mathit{P}_1^{q_1+1}$ is decoded which is incorrect. Here, $ \mathit{P}_2^{2^{\eta_2}}$ is chosen, as from \eqref{constraint_3}, this is the maximum value of the constellation points of Tx2. The condition for zero BER in ideal conditions considering $L$ Txs is given as follows
\begin{equation}
 \mathit{P_x}^{q_x}\mathit{h_x}+ \sum_{r=1}^{\mathit{L-x}} \max_{\forall q_{x+r} \in \mathbf{Q}_{x+r}} \{ \mathit{P}_{x+r}^{q_{x+r}}\mathit{h}_{x+r}\} < \frac{\mathit{P_x}^{q_x+1}\mathit{h_x}+\mathit{P_x}^{q_x}\mathit{h_x}}{2}.
 \label{eq_1}
\end{equation}
From \eqref{constraint_3},
$
\max_{\forall q_{x+r} \in \mathbf{Q}_{x+r}} \{\mathit{P}_{x+r}^{q_{x+r}}\mathit{h}_{x+r}\}=\mathit{P}_{x+r}^{2^{\eta_{x+r}}}\mathit{h}_{x+r}.
$
Hence, \eqref{eq_1} can be written as follows
\begin{equation}
 \mathit{P_x}^{q_x}\mathit{h_x}+ \sum_{r=1}^{\mathit{L-x}}\mathit{P}_{x+r}^{2^{\eta_{x+r}}}\mathit{h}_{x+r} < \frac{\mathit{P_x}^{q_x+1}\mathit{h_x}+\mathit{P_x}^{q_x}\mathit{h_x}}{2}.
\label{eq_2}
\end{equation} 
On further simplification, \eqref{eq_2} becomes
\begin{equation}
  \mathit{P_x}^{q_x+1} >  \frac{2 }{\mathit{h_x}} \sum_{r=1}^{\mathit{L-x}}  \mathit{P}_{x+r}^{2^{\eta_{x+r}}} \mathit{h}_{x+r}+\mathit{P_x}^{q_x}.
\label{eq_3_1}
\end{equation}
Let 
\begin{equation}
\Delta_x^{q_x+1} = \frac{2 }{\mathit{h_x}} \sum_{r=1}^{\mathit{L-x}}  \mathit{P}_{x+r}^{2^{\eta_{x+r}}} \mathit{h}_{x+r}+\mathit{P_x}^{q_x}.
\label{for_algo}
\end{equation}
Substituting \eqref{for_algo} in \eqref{eq_3_1}, we get
\begin{equation}
\mathit{P_x}^{q_x+1} > \Delta_x^{q_x+1}.
\label{eq_4}
\end{equation}
Finally, \eqref{eq_4} can be rewritten as follows
$$
\mathit{P_x}^{q_x+1}=
\begin{cases}
\Delta_x^{q_x+1}+1 \,\,\mbox{if}\,\, \Delta_x^{q_x+1} \in \mathbb{Z^+} \,,\\
\left\lceil \Delta_x^{q_x+1} \right\rceil \mbox{otherwise}.
\end{cases}
 = \left\lfloor \Delta_x^{q_x+1}+1 \right\rfloor.
$$
Next, we consider Algo.~\ref{algo_1} to obtain constellation points for $L$ Txs with desired spectral efficiency for each of the Txs.
The spectral efficiency of the VLC system with $L$ Txs with the proposed scheme is $\eta_1+\eta_2+\ldots+\eta_L=\sum_{x=1}^{L}\eta_x$ bpcu.
\begin{table}[t]
\renewcommand{\arraystretch}{1.3}
\begin{center}
		\caption{Decoding computational complexity of MAC.}
		\label{comp_comp}\begin{tabular}{ |c||c| }
\hline
Decoding scheme   & number of ML computations \\\hline \hline
JML decoding & $\prod_{x=1}^{L}2^{\eta_x}$   \\\hline
SIC based decoding&$\sum_{x=1}^{L}2^{\eta_x}$   \\\hline
$M$ JML + $(L-M)$ SIC decoding& $\prod_{x=1}^{M}2^{\eta_x}+\sum_{x=M+1}^{L}2^{\eta_x}$ \\\hline
\end{tabular} 
\end{center}
\end{table}
\subsubsection{Decoding Mechanism}
Decoding for the proposed scheme can be performed using SIC decoding or JML decoding. In SIC based decoding, decoding of the signal from the $x^{th}$ Tx happens after performing SIC of the signals received from $1^{st}$ to $(x-1)^{th}$ Tx and the received signals from $(x+1)^{th}$ to $L^{th}$ Tx is treated as noise as given in Algo.~\ref{SIC_decoding}. The decoding order of the Txs data at the Rx is given as $\mathcal{D}_{\mbox{Tx1}} < \mathcal{D}_{\mbox{Tx2}} < \ldots < \mathcal{D}_{\mbox{TxL}}$, where $\mathcal{D}_{\mbox{Tx}x}$ denotes the SIC decoding order of $x^{th}$ Tx. We assume that the constellation points corresponding to all the Txs are known at the Rx. In Algo.~\ref{SIC_decoding}, $\hat{\mathit{P}_x}$ denotes the power estimate of $x^{th}$ Tx. 
In JML decoding, we perform ML decoding over all possible combinations of constellation points to decode as follows
\begin{equation*}
 [ \hat{\mathit{P}}_1 \hat{\mathit{P}}_2 \ldots \hat{\mathit{P}_L}] = \mbox{min}_{\forall q_x \in \mathbf{Q}_x} ||y-\sum_{x=1}^{L}h_x \widetilde{\mathit{P}_x^{q_x}} ||.
\end{equation*}
In  $M$ JML + $(L-M)$ SIC decoding, the data from $M$ Txs is decoded using JML decoding and the data from $(L-M)$ Txs is decoded using SIC for $M\geq 2 \,\, \mbox{and}\,\, M<L$. Note that when $M=0$, the $M$ JML + $(L-M)$ SIC decoding is same as SIC decoding, and when $M=L$, it becomes JML decoding.
The number of computations with JML decoding, SIC based decoding, and $M$ JML + $(L-M)$ SIC decoding is given in Table~\ref{comp_comp}.
It can be observed that as the value of $L$ and $\eta_x$ increase, the computations involved in JML decoding will be significantly higher as compared to SIC based decoding. Further, for $M$ JML + $(L-M)$ SIC decoding, the number of computations also depend on $M$.
\begin{figure}[!t]
\centering
\includegraphics[width=3.4in,height=1.25in]{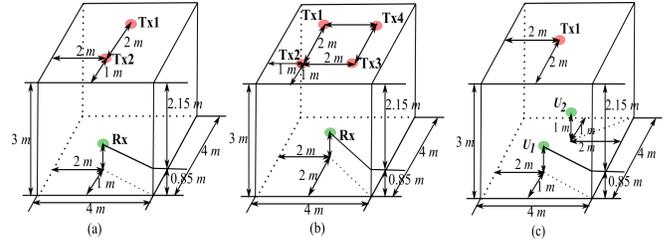}
\caption{Room showing the position of (a). 2 Txs and Rx in MAC (b). 4 Txs and Rx in MAC (c). Tx and 2 users in BC.}
\label{room_scenario_BC}
\end{figure}
Given this, for any value of $L$, the optimal value of $M$ denoted by $\hat{M}$ that minimizes the computations and achieves the desired BER for any $x^{th}$ Tx is computed as follows
\begin{equation*}
\hat{M} = \min_{M}\left\lbrace \prod_{x=1}^{M}2^{\eta_x}+\sum_{x=M+1}^{L}2^{\eta_x} \right\rbrace,\ s.t., \,\, 0 \leq M \leq L,
\label{optimal M}
\end{equation*}
\begin{equation}
\mbox{BER of}\,\, x^{th}\,\, \mbox{Tx} \leq 10^{-v},\,\, v \in \mathbb{Z^+},\, \mbox{and}\,\,
\widetilde{SNR} \leq \Gamma \, dB,
\label{optimal M1}
\end{equation}
where, $\widetilde{SNR}$ is the average received signal-to-noise ratio ($SNR$) and $\mbox{ BER \,\,of}\,\, x^{th}\,\, \mbox{Tx}$ imply the minimum BER to be achieved by the $x^{th}$ Tx.
The BER constraint in \eqref{optimal M1} together with \eqref{scenario1_constraint} ensures that all the Txs from 1 to $x-1$ also achieve this BER as the Txs till $x-1$ have lower decoding order/higher received power.
Assuming same spectral efficiency at system level between OMA and NOMA, the average number of ML computations involved in JML decoding ($L 2^{\sum_{i=1}^{L}\eta_i}$) is never greater than the computations involved in OMA. We can prove this by assuming the average number of ML computations in JML decoding is less than or equal to the computations involved in OMA ($\sum_{i=1}^{L}2^{L \eta_i}$) and then showing the assumption is indeed correct as follows
\begin{equation}
L 2^{\sum_{i=1}^{L}\eta_i} \leq \sum_{i=1}^{L}2^{L \eta_i}.
\label{proof_1}
\end{equation}
On simplification of \eqref{proof_1} using $\mbox{log}(\theta)<\theta\, \forall\, \theta>0$  gives
\begin{equation*}
\sum_{i=1}^{L}\eta_i-\frac{1}{L}\sum_{i=1}^{L}\eta_i \geq 1 \, \forall \, L \geq 2.
\label{proof_final}
\end{equation*}
The above condition is always true for $L \geq 2$ and $\eta_i\geq 1\,\forall\, i\,\in\{1,\,2,\,\ldots,L \}$ and hence the proof.
\begin{table}[t]
\renewcommand{\arraystretch}{1.15}
\begin{center}
		\caption{Parameters for simulation \cite{intro_1_2}.}
		\label{sim_params}\begin{tabular}{ |c||c| }
\hline
Parameter   & Value \\\hline \hline
Dimensions of the room (length $\times$ breadth $\times$ height) & $4 \times 4 \times 3 \, m^3$   \\\hline
LED semi angle, $\Phi_{1/2}$ & $60^\circ$ \\\hline
PD FOV, $\psi_{fov}$ & $60^\circ$ \\\hline
PD responsivity, $R_p$ & 0.4 $A/W$ \\\hline
PD detection area, $A_D$ & $10^{-4} \,\, m^2$ \\\hline
Refractive index of optical concentrator at Rx, $\eta$& 1.5\\\hline
Optical filter gain, $T$ &  1 \\\hline
\end{tabular} 
\end{center}
\end{table}
\subsection{BC}
Without loss of generality, here we assume $K$ users and each user is equipped with a single PD Rx. The constellation points of users are similar to that of the constellation points of the Txs in MAC. Similar to MAC, the constraint on the power per channel use by the Tx is given as $\sum_{\alpha=1}^{K} \widetilde{\mathit{U}_\alpha^{2^{\eta_\alpha}}} \leq \mathcal{Q}$, where $\mathcal{Q}$ is the available transmit power per channel use and $\widetilde{\mathit{U}_\alpha^{2^{\eta_\alpha}}}$ is the normalized constellation point of the $\alpha^{th}$ user. The data is decoded at the users' end by employing SIC. The received signal at $\alpha^{th}$ user is given as
$y_{\alpha}=S_{\alpha}+n_{\alpha},$
where, $n_\alpha$ is the 0 mean real valued AWGN at the $\alpha^{th}$ user with $\sigma_\alpha^2$ variance and
$S_{\alpha}$ is the received signal at the $\alpha^{th}$ user. The $S_{\alpha} = \left( \widetilde{\mathit{U}_1^i}+\widetilde{\mathit{U}_2^j}+\ldots+\widetilde{\mathit{U}_K^w}\right)g_{\alpha}$, where, $g_\alpha$ is the channel gain between Tx and  $\alpha^{th}$ user. Here, $\widetilde{\mathit{U}_\alpha^r}$ denotes the transmit power corresponding to $\alpha^{th}$ user for $\alpha \in \{1,2, \ldots, K \}$ and $r \in \{1,2, \dots, 2^{\eta_\alpha} \}$. Similar to MAC, the decoding order of users in BC is given as $\mathcal{D}_{U_1} < \mathcal{D}_{U_2} < \ldots < \mathcal{D}_{U_K}$, where, $\mathcal{D}_{U_\alpha}$ denotes the SIC decoding order of the $\alpha^{th}$ user.  The number of ML computations in decoding $\alpha^{th}$ user is $\sum_{i=1}^{\alpha} 2^{\eta_i}$. This value includes the ML computations involved in SIC process of $\alpha-1$ users while decoding $\alpha^{th}$ user's data.
The computational complexity of the proposed NOMA, DCO-OFDM NOMA, and OMA is compared in Table~\ref{com_complexity} for 2 users in BC. We consider split radix FFT \cite{srfft} is used in DCO-OFDM NOMA which is one of the known computationally efficient FFT algorithms. It is observed that the computational complexity involved in proposed NOMA is significantly less as compared to DCO-OFDM NOMA and OMA.
\section{Numerical Results}
We present numerical results for the scenarios in Fig.~\ref{room_scenario_BC} using the parameters in Table~\ref{sim_params}.
For MAC, $\widetilde{SNR}$ is computed as follows
\begin{equation*}
\widetilde{SNR} = \mathop{\mathlarger{\mathlarger{\mathbb{E}}}}_{\forall q_x \in\mathbf{Q}_x}\left( \frac{\left(\sum_{x=1}^L\widetilde{P_x^{q_x}}h_x\right)^2 }{\sigma^2}\right).
\end{equation*}
Similarly, for BC, the received $SNR$ at $\alpha^{th}$ user denoted by $\widetilde{SNR_\alpha}$ is computed as follows
\begin{equation*}
\widetilde{SNR_\alpha}= \mathop{\mathlarger{\mathlarger{\mathbb{E}}}}_{\forall j \in \{1,\,2,\,\ldots,2^{\eta_k} \} } \left( \frac{\left(\sum_{i=1}^{K} \widetilde{U_{i}^j}\right)^2h_\alpha^2}{\sigma_{\alpha}^2}\right),\, k \in \, \{1,\,2,\,\ldots,K \}.
\end{equation*}
We assume $\mathcal{P}$ = $\mathcal{Q}$ = 1 for the presented simulation results, where BER is numerically evaluated using Monte Carlo simulations.
\begin{figure}[!t]
\centering
\includegraphics[width=3.05in,height=1.8in]{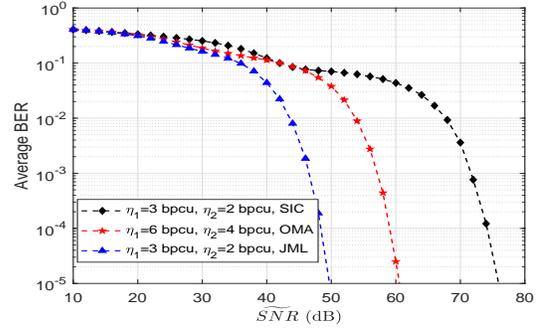}
\caption{BER of proposed scheme with SIC decoding, JML decoding, and OMA with 2 Txs in MAC.}
\label{ber1_scenario1_2tx}
\end{figure}
\begin{figure}[!t]
\centering
\includegraphics[width=3.05in,height=1.8in]{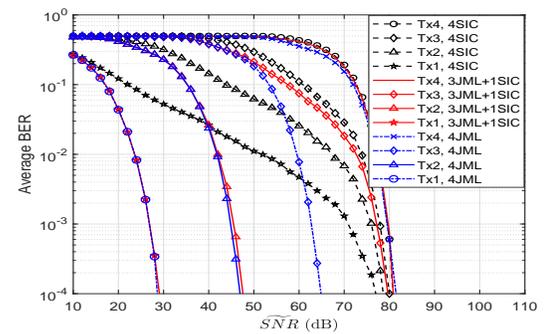}
\caption{BER of proposed scheme with JML, SIC, and $M$ JML + $(L-M)$ SIC decoding for $M=0,\,\,3,\,\, \mbox{and} \,\,4$ with 4 Txs in MAC with $\eta_1=\eta_2=\eta_3=\eta_4=$ 2 bpcu.}
\label{ber1_scenario1}
\end{figure}
For MAC, we define average BER as the average of the BER of all the Txs using same decoding mechanism.
In Fig.~\ref{ber1_scenario1_2tx}, the average BER using JML decoding and SIC decoding is compared for the scenario shown in Fig.~\ref{room_scenario_BC} (a). It is observed that the BER with JML decoding is significantly better as compared to the SIC decoding. However, this improved performance with JML decoding comes at the cost of computational complexity. It is also observed that the proposed scheme with JML decoding performs better as compared to OMA \cite{oma_res}, with less computational complexity.

\begin{table*}[t]
\renewcommand{\arraystretch}{1.25}
\begin{center}
		\caption{Comparison of computational complexity for 2 users in BC.}
		\label{com_complexity}\begin{tabular}{ |c||c|c|c|c|c|c| }
\hline
\multirow{3}{*}{Computation type}   & \multicolumn{3}{c|}{for arbitrary spectral efficiency and N-FFT}& \multicolumn{3}{c|}{for $\eta_1=\eta_2=7\, bpcu, \mbox{N}=256 $ FFT}\\
\cline{2-7}
  & proposed & DCO-OFDM  & \multirow{2}{*}{OMA} & proposed  & DCO-OFDM & \multirow{2}{*}{OMA} \\
    &  NOMA & NOMA \cite{ofdm_1} &  &  NOMA &  NOMA \cite{ofdm_1} &   \\\hline \hline
ML computations & $\sum_{i=1}^{2}\sum_{j=1}^{i}2^{\eta_j}$ &$\sum_{i=1}^{2}\sum_{j=1}^{i}2^{\eta_j}$ & 0.5$\sum_{i=1}^{2}2^{2\eta_i}$ &384 &384 & 16384\\\hline
Real multiplications at FFT \& IFFT&- & 5$\left( \mbox{N} \mbox{log}_2 \mbox{N} - 3\mbox{N} +4 \right)$&- & - & 6420 &-\\\hline

Real additions at FFT \& IFFT&- & 5$\left( 3\mbox{N} \mbox{log}_2 \mbox{N} - 3\mbox{N} +4 \right)$&- & -&26900&- \\\hline

DC bias addition/removal&- & 3$  \mbox{N}$ &-& - & 768&- \\\hline
\end{tabular} 
\end{center}
\end{table*}

In Fig.~\ref{ber1_scenario1},  the BER using JML decoding, SIC decoding, and $M$ JML + $(L-M)$ SIC decoding for $M=0,\,\,3,\,\,\mbox{and}\,\,4$ is shown for the scenario in Fig.~\ref{room_scenario_BC} (b). 
It is observed that the BER of JML decoding and $M$ JML + $(L-M)$ SIC decoding is same for $M-1$ Txs since the most interfering Tx till $M-1$ Txs are decoded using JML decoding. It is also observed that the BER depends on the value of $M$ and the improved performance is achieved at the cost of increased computations.
\begin{table}[t]
\renewcommand{\arraystretch}{1.10}
\begin{center}
		\caption{The $\hat{M}$ for $\Gamma=70$, $L=4$, and $\eta_x = 2 $ bpcu $\forall\,x\,\in\{1,\,2,\,3,\,4\}$ for the scenario in Fig.~\ref{room_scenario_BC} (b).}
		\label{BER-control}\begin{tabular}{ |c|c|c|c|c| }
\hline
$v$ in    & $\hat{M}$ for BER  & computations & $\hat{M}$ for BER  & computations \\
\eqref{optimal M1}  & constraint  & at Rx &  constraint  &at Rx \\
 & on Tx2 &  & on Tx1 &   \\\hline \hline
2 & 0&16 &0 & 16 \\\hline
3& 3  &68 &2 &24 \\\hline
6& 3 &68 & 2& 24\\\hline
8& 3 &68 & 2& 24\\\hline
\end{tabular} 
\end{center}
\end{table}
In Table~\ref{BER-control}, the value of $\hat{M}$ in \eqref{optimal M1} is numerically evaluated and it is observed that as the BER constraint on lower power allocated Txs increase, the computational complexity increases since the value of $\hat{M}$ increases faster for lower power allocated Txs as compared to higher power allocated Txs.

\begin{figure}[!t]
\centering
\includegraphics[width=3.05in,height=1.9in]{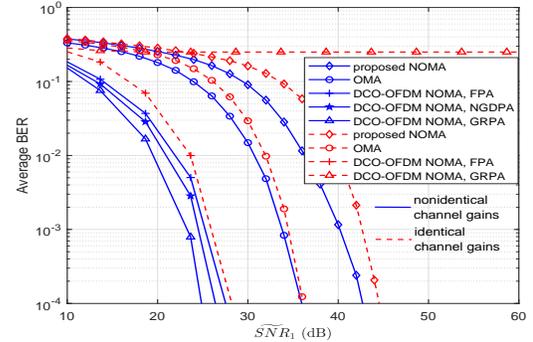}
\caption{BER of proposed NOMA, OMA, and DCO-OFDM NOMA with 2 users in BC, where $\eta_1=\eta_2=2$ bpcu for proposed NOMA, DCO-OFDM NOMA and $\eta_1=\eta_2=4$ bpcu for OMA.}
\label{averageBER_BC}
\end{figure}

In Fig.~\ref{averageBER_BC}, the average BER of the proposed NOMA, OMA, and DCO-OFDM NOMA with fixed power allocation (FPA) for a power allocation coefficient of $2/3$, gain ratio power allocation (GRPA) \cite{intro_3_1}, and normalized gain difference power allocation (NGDPA) \cite{ofdm_2} is compared for a 2 user BC for two cases namely nonidentical channel gains as shown in Fig.~\ref{room_scenario_BC} (c) and identical channel gains, where $U_2$ is assumed to be at the same position as $U_1$ in Fig.~\ref{room_scenario_BC} (c). We considered $\eta_1=\eta_2=2$ bpcu for proposed NOMA, DCO-OFDM NOMA, and $\eta_1=\eta_2=4$ bpcu for OMA for a fair comparison.
It is observed that proposed NOMA performs inferior as compared to OMA and DCO-OFDM NOMA in terms of BER. However, the improved performance with OMA/DCO-OFDM NOMA comes at the cost of increased computational complexity as shown in Table~\ref{com_complexity}. Since high channel correlations is a common issue in indoor VLC, we have analysed a scenario where channel gains are identical. For such a scenario, it is observed that the performance of DCO-OFDM NOMA with GRPA is severely degraded compared to FPA. Note that for perfectly identical channel gains, NGDPA is not applicable as power allocated to one of the users becomes zero.
Further, the implementation of proposed scheme is simpler as compared to DCO-OFDM NOMA and also the latency experienced in DCO-OFDM NOMA would be higher due to it's complex system model involving IFFT and FFT at Tx and Rx, respectively.

\section{Conclusions}
In this letter, we have proposed a generalized scheme for MAC and evaluated its performance in terms of BER and computational complexity with SIC based decoding, JML decoding, and $M$ JML + $(L-M)$ SIC decoding. It is observed that the gain in BER with JML decoding comes at the cost of computations and the performance of $M$ JML + $(L-M)$ SIC decoding depends on $M$.
For MAC, it is observed that the proposed scheme with JML decoding performs better than the state-of-the-art OMA both in terms of BER and computations. 
For BC, the BER performance of proposed scheme is inferior as compared to OMA and DCO-OFDM NOMA. However, the proposed scheme outperforms both the existing schemes in terms of computational complexity, and hence, the proposed scheme is suitable for high SNR regimes providing low computational and system complexity.

\end{document}